**Optimal design for wavelength conversion with a configuration of walk-off compensation in free space in the nanosecond pulsed regime**

Chiaki Ohae,[1,2] Kenji Hasegawa,[2] Masato Nagano,[2] Soma Tahara,[2] and Masayuki Katsuragawa[1,2,*]

[1]*Institute for Advanced Science, University of Electro-Communications, 1-5-1 Chofugaoka, Chofu, Tokyo 182-8585, Japan*
[2]*Graduate School of Informatics and Engineering, University of Electro-Communications, 1-5-1 Chofugaoka, Chofu, Tokyo 182-8585, Japan*
[*]E-mail: katsuragawa@uec.ac.jp

**Abstract**

On the basis of standard wavelength conversion by the use of angular phase matching of nonlinear optical crystals arranged in free space, applicable to a wide range of wavelengths and laser intensities, we both numerically and experimentally present an optimal design for achieving near-full energy conversion while maintaining good single-mode properties of fundamental laser radiation in the nanosecond regime.

A single-frequency nanosecond pulsed laser (with a tunable wavelength) is a unique light source that has a high peak intensity (typically at a magnitude of MW), transform-limited frequency purity (typically at tens of MHz) [1−3], and a single transverse mode property. It becomes a particularly powerful tool in studies of high-resolution "nonlinear" laser spectroscopy [4,5]. As in many cases, the method of wavelength conversion using nonlinear optical crystals is an efficacious way to extend the available wavelength range of such a laser source. However, especially in such applications as the above, it is critical to realize not only a high conversion efficiency but also a high beam quality maintaining the original single-mode properties of the light source. If we apply the widely used method of angular phase matching in nonlinear optical crystals [6,7] in the nanosecond regime, as is known well, it causes a serious difficulty: beam quality degradation due to the walk-off, which inevitably appears in the conversion process, limiting, in turn, the conversion efficiency. This is mainly due to the intrinsic property of wavelength conversion in the nanosecond regime that the peak intensities of nanosecond pulses are generally fairy lower than those of picosecond or femtosecond laser sources, and thereby require a longer crystal length to obtain a high conversion efficiency.

The recently developed method, non-critical phasematching [8,9] or quasi-phasematching represented by periodically poled lithium niobate [10], has emerged as a technology which avoids the walk-off problem and thereby achieves a high conversion efficiency while maintaining the high beam quality of the original laser radiation. However, at present they do not necessarily provide good solutions for a variety of wavelength conversions in the nanosecond pulsed regime, since the applicable wavelengths are limited, and in the latter case, scaling up to a high-energy regime is also limited owing to crystal damage problems (as the crystal cross-sectional area is technologically limited).

In this study, we describe how we can overcome these problems on the basis of the standard angular phase matching method, which is widely applicable with respect to wavelength, laser intensity, and other factors. We theoretically and experimentally show a quantitative design by using a series of nonlinear optical crystals arranged in free space, which achieves a near-full energy conversion efficiency while maintaining good single mode properties of the fundamental laser source.

Before proceeding with main description, we briefly comment on closely related works. Conversion efficiencies of 83% (second harmonic generations of 1.064-μm nanosecond YAG lasers) have been reported by using a hat-top beam profile with a uniform cross-sectional intensity distribution [11,12]. Although this approach is suitable for achieving a high conversion efficiency for laser beams with multiple transverse modes, it is not necessarily suitable for applications where a single transverse mode property plays an essential role in phenomena such as those in nonlinear optical processes. Also, nonlinear wavelength conversions with a configuration of a walk-off compensate scheme, described in this study, have been achieved by using an optical contact technique [13−15] or a room-temperature bonding technology pioneered by Shoji et al. [16,17]. Although these are excellent methods, they currently have a definite limitation in cases where it is technically difficult to precisely cut a crystal-axis angle out with an accuracy that satisfies the phase matching condition, or where these methods must be used for wavelength tunable lasers. The method described in this study is advantageous in that it can be broadly applied including these cases.

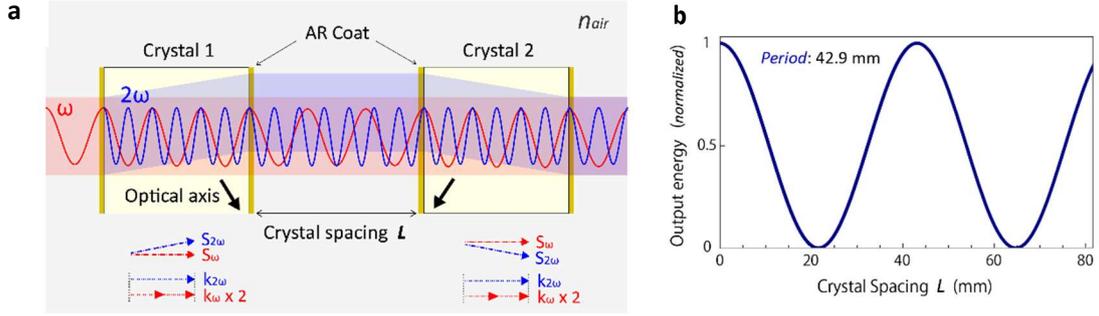

**Fig. 1 Second harmonic generation by a pair of crystals arranged in a walk-off-compensate configuration.**
**a**, Basic configuration. **b**, Behavior of second harmonic generation energy versus spacing between crystals.

Here, we describe second harmonic generation as an example. Figure 1a illustrates a basic configuration of the method. To compensate for degradation of the transverse mode and conversion efficiency by the walk-off, we place a pair of nonlinear crystals with their optical axes inverted to each other (Fig. 1a) [18–20]. Here, the second harmonics generated in each crystal must overlap in phase so that we do not lose the effect of using the crystal pair. Usually, the front and end faces of the crystals are anti-reflection coated (for the fundamental and second harmonics) and are separated by a gaseous medium and windows. If we use such a setup that the nonlinear crystals are placed in air, the phase difference $\Delta\phi$, which arises between the second harmonics generated in the two crystals, is given by Eq. (1):

$$\Delta\phi = \left(\frac{2\pi n_2}{\lambda_2} - 2\frac{2\pi n_1}{\lambda_1}\right) \times L + 2\Delta\phi_{AR} \qquad (1)$$

where the subscripts 1 and 2 mean fundamental and second harmonics, $\lambda$ is the wavelength, $n$ is the refractive index of atmosphere depending on the wavelength [21], $L$ is the distance between crystals, and $\Delta\phi_{AR}$ is the phase difference generated by the AR coat at the crystal end and front faces [22, 23]. To superimpose the second harmonics generated by each of the two crystals in phase, this $\Delta\phi$ must be an integer multiple of 2π. Manipulating the crystal spacing, $L$, as an adjusting parameter is one solution to satisfy such a condition.

Figure 1b shows the calculated second harmonic generation (375 nm) when the phase matching condition is satisfied, where lithium triborate (LBO, type-I, fundamental: ordinary polarization, second harmonic: extra-ordinary polarization) crystals are used as nonlinear optical crystals and the fundamental wavelength is set to 750 nm (phase matching angle 37.15°, walk-off angle 1.032°). The crystal spacing, giving $\Delta\phi$ = 2π, amounts to $L$ = 42.9 mm at 21 °C and 1 atm [21], and the second harmonic generation (normalized by the peak intensity) oscillates sinusoidally with this spacing as a period (Fig. 1b). The effect of the AR coatings is excluded here because it may only shift the origin of the horizontal axis in Fig. 1b.

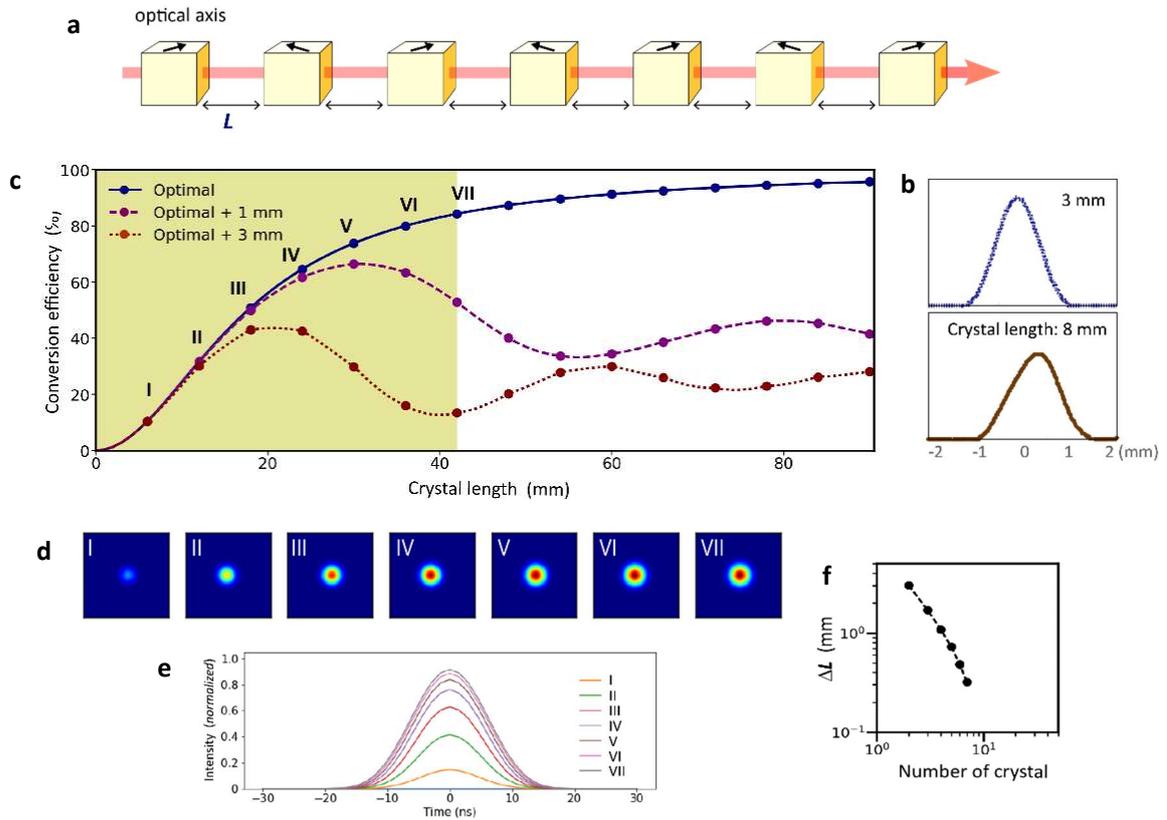

**Fig. 2. Second harmonic generation by a LBO crystal pair series (numerical calculation).** **a**, Configuration of a crystal pair series. **b**, Beam profiles for crystal lengths of 3 and 8 mm. **c**, Energy conversion efficiency versus number of crystals. Three cases are shown (see figure key): optimal crystal-spacing, +1 mm deviation, and +3 mm deviation. **d**, Beam profiles at crystal exits I to VII (in optimal configuration). **e**, Temporal profiles at crystal exits I to VII (in optimal configuration). **f**, Accuracy of positioning of the crystals, required for keeping the conversion efficiency reduction within 5% for the achievable maximum efficiency ("optimal").

The subject of this study is to theoretically and experimentally clarify a design that can achieve a highly efficient wavelength conversion while maintaining single mode properties, on the basis of the fundamental configuration illustrated in Fig. 1a.

If compensating the walk-off is the only consideration, it is best to divide the crystal length as short as possible and to place a large number of crystals in the inverted configuration as shown in Fig. 1a and provide a sufficient crystal length for the near-full conversion. In reality, however, practical problems appear; they include unavoidable residual reflection losses at the crystal front/end surfaces and also an installation problem how, within the Rayleigh length, we may arrange a large number of crystals including mechanisms so as to adjust the phase matching angle and the crystal spacing. Therefore, a practically optimal design means to realize our purpose by the fewest divisions of crystal which has a sufficient length for near-full conversion. As shown in Fig. 2a, we arranged a series of LBO crystal pairs with the reversed optical axis layout shown in Fig. 1a, and then numerically investigated the conditions for maintaining the single mode properties up to the full conversion regime under a variety of crystal lengths, beam diameters, and laser intensities.

As one of the substantial conclusions, it was clarified that the main criterion for maintaining a single transverse mode was to limit the walk-off to <1/10 of the fundamental beam diameter at $1/e^2$ during the wavelength conversion process. Figure 2b shows cases where the walk-off meets this criterion (upper panel, crystal length 3 mm) or exceeds it (lower panel, crystal length 8 mm). It was also confirmed for minimizing the number of dividing the crystal that the beam diameter should be set to the narrowest so as not to damage the crystals. This is due to an intrinsic mechanism by which the beam shift caused by the walk-off becomes relatively larger in relation to the beam diameter, namely, the required number of dividing the crystal is increased in relation to the beam diameter, as the crystal length required for the near-full conversion is proportionally longer to square of the beam diameter.

Figure 2c presents a result numerically calculated by considering these requirements. The fundamental laser beam was assumed to be a fundamental Gaussian beam with an intensity of 175 $MW/cm^2$ (pulsed energy 10.6 mJ), a temporal duration of 16.5 ns at full width at half maximum (FWHM), a transform-limited linewidth of 20 MHz at FWHM, and beam diameters at $1/e^2$ of 1.22 mm (horizontal direction) and 1.15 mm (vertical direction), corresponding to a Rayleigh length of 1300 mm. The crystal length was set to 6 mm arousing a walk-off corresponding to 9% of the beam diameter (1.2 mm), close to the optimal length. The crystal spacing, $L$, was assumed to be exactly adjusted to the optimal length of 42.9 mm as confirmed in Fig. 1b. The second harmonic generation (dark blue circles) approached full conversion under these conditions (Fig. 2c), while maintaining a single transverse mode (Fig. 2d) and a smooth pulsed envelope (Fig. 2e). The energy conversion efficiency reached 82% at the 7th crystal exit (96% at the 15th exit).

The accuracy required for the crystal spacing is also an another critical factor for use of this method in reality. Figure 2c shows cases where the crystal spacing deviated from the optimal length by +1 mm (purple circles) and +3 mm (brown circles). As expected, the larger the deviation was, the lower the conversion efficiency reached, and then it decayed with a slowly oscillating structure. This placement of the crystals requires higher accuracy as the total crystal (interaction) length is longer. The mechanism is physically analogous to that in the phase matching condition, in which the required accuracy is inversely proportional to the total crystal length. In fact, plotting the deviations from the optimal crystal spacing, $\Delta L$, that reduced the conversion efficiency by 5% from that of the optimal configuration at each crystal exit (Fig. 2c, dark blue circles) confirmed that the behavior of $\Delta L$ was approximately inversely proportional to the total crystal length (Fig. 2f). This calculated result clarifies that each crystal must be placed with a high positioning accuracy of <0.3 mm (typically 0.1 mm) in the case of seven crystals (total interaction length = 42 mm) to enable an energy conversion efficiency of 82%.

The effects of the temperature and pressure of the air in which the crystals are arranged should also be taken into account when the crystal spacing is determined in reality. The above results, however, show that the required control accuracies to be <2 °C and <750 Pa, namely, this point does not technically become a major obstacle.

In the crystal arrangement in Fig. 2a, the optical axis of one of the crystal pair can be optionally inverted in the horizontal plane, as in the quasi-phase-matching method. This optional arrangement shifts the optimal crystal spacing by half a period (21.45 mm). If the crystal spacing is longer than the half period, this option can be used to shorten the crystal spacing and thereby the entire crystal length too. As all crystals must be placed within the Rayleigh length of the fundamental and the second harmonic beams, this option of shortening the total crystal length is effective.

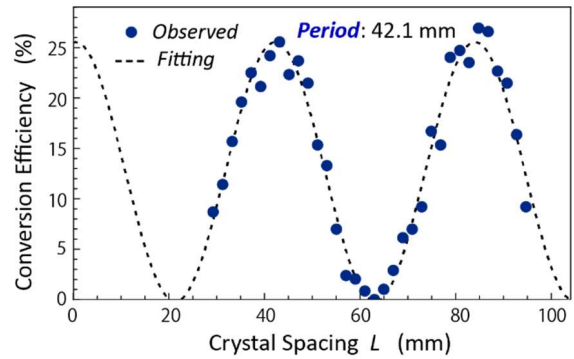

**Fig. 3. Fundamental property of second harmonic generation by a pair of crystals: energy conversion efficiency versus crystal spacing.**

It is not obvious whether we can obtain the results in reality as predicted in the numerical calculation. Below, we show an experimental demonstration examined according to the predictions in the numerical calculation.

As the fundamental laser, we used an injection-locked nanosecond pulsed Ti sapphire laser [1–3] (single transverse mode, $M^2$ = 1.1; spectral width, 20 MHz). All of the experimental conditions were set to the same parameters as used in the numerical calculations. Also, we used a power meter (Gentec, QE25LP) to measure the energy, a CCD-based beam profiler (Gentec, Wincam D) to measure the beam profile, and a fast photodetector (Hamamatsu Photonics K.K., R1328U, buildup time 60 ps) and a fast digital oscilloscope (Tektronix, DPO 7254, bandwidth: 2.5 GHz) to measure the temporal waveform.

Figure 3 shows the second harmonic energies (dark blue circles) obtained by using a pair of LBO crystals (crystal length 6 mm) at different crystal spacings. First, the phase matching angle was adjusted with a resolution of 0.05 degree, and then, the crystal spacing was varied. The periodic behavior of the generated second harmonic energies (period, 42.1 mm) was observed, in good agreement with the prediction (period, 42.9 mm; Fig. 1b). The variation of the generated second harmonic energies was due mainly to the energy fluctuation of the fundamental laser. The slight shift of the origin in the periodic behavior was due to the refractive index dispersion by the anti-reflection coatings on the front and end faces of the crystals.

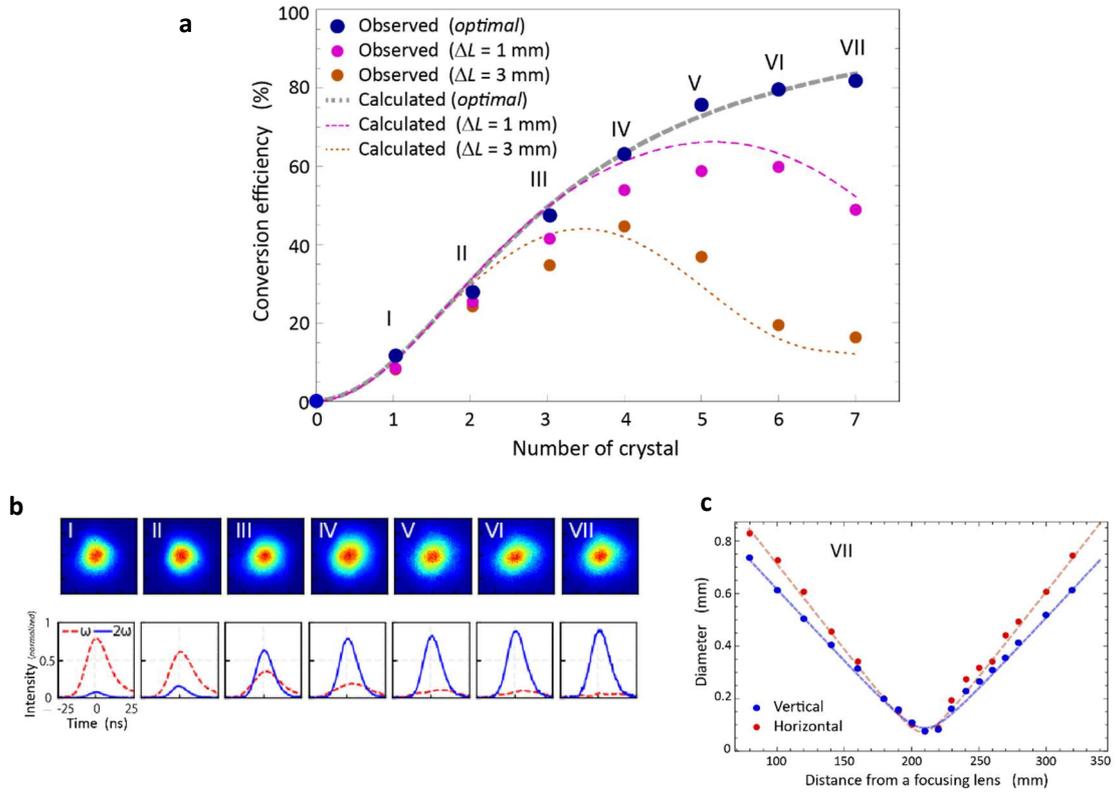

**Fig. 4. Second harmonic generation by a crystal pair series (maximum of 7 crystals). a**, Energy conversion efficiency versus number of crystals. Three cases are shown: optimal crystal spacing (dark blue circles), and deviations of +1 mm (purple circles) and +3 mm (brown circles). **b**, Beam profiles (upper panel) and temporal waveforms (lower panel; red dotted line, fundamental; blue solid line, second harmonic) at each of the crystal exits I to VII, with the optimal configuration. **c**, Estimation of beam quality of the second harmonic with a maximal conversion efficiency at VII, based on the $M^2$ method.

As we confirmed the fundamental performance of the proposed scheme, we assessed second harmonic generation based on the configuration of a series of seven LBO crystals (6 mm each). We accurately placed the seven LBO crystals one by one at the optimal spacing and measured the characteristic behaviors of the generated second harmonics. The adjustment procedure was same as that employed in Fig. 3. The total length of the series of crystals was 293 mm, enough shorter than the Rayleigh length (1300 mm for the fundamental). Thus, the crystals were placed within the range regarded as a collimated beam.

Figure 4a plots the measured second harmonic energy conversion efficiencies (dark blue circles) against the total numbers of crystals placed. The properties well reproduced the numerically predicted result at the optimal crystal spacing (Fig. 4a, black dashed line). The energy conversion efficiency reached 82% (generated energy, 8.7 mJ) at the exit of the seventh crystal (total crystal length, 42 mm).

The pink and brown circles show the case giving deviations of +1 mm and +3 mm from the optimal crystal spacing, respectively. Although the deviations were slight, the conversion efficiencies were seriously reduced from the optimal as predicted (Fig. 2c, purple and brown dashed lines) and reached saturation at 60% (at the exit of the 6th crystal) for the +1-mm deviation case and 45% (at the exit of the 4th crystal) for the +3-mm deviation case. The importance of accumulating the generated second harmonics accurately in phase to achieve a near-full conversion efficiency was also recognized based on experiment.

Figure 4b shows the beam profiles (upper panel) of the generated second harmonics and the temporal waveforms of the fundamentals (lower panel, red dashed line) and the second harmonics (lower panel, blue solid line), measured at the optimal crystal spacing. The beam profiles of the second harmonics were maintained consistently with a smooth intensity distribution over the entire wavelength conversion process and reached the seventh crystal exit. The beam diameters were measured as 1.2 mm at $1/e^2$ (walk-off direction) and 1.1 mm (vertical direction). The temporal waveforms of the second harmonics (pulse duration at the seventh crystal exit, 14.8 ns at FWHM) were also maintained with a smooth pulsed envelope reflecting the incident fundamental waveform.

Finally, by applying the $M^2$ method, we quantitatively evaluated the beam quality of the achieved second harmonic at the seventh crystal exit (Fig. 4c), using a lens with a focal length of 200 mm. As confirmed in Fig. 4c, the $TEM_{00}$ single transverse-mode property ($M^2$ = 1.0) was indeed achieved. The slightly elliptical beam shape was due to that of the fundamental laser employed.

**Conclusion**

On the basis of standard wavelength conversion by the use of angular phase matching in nonlinear optical crystals arranged in free space, which can be applied to a wide range of wavelengths and laser intensities, we have both numerically and experimentally shown the optimal design for achieving near-full conversion efficiency while maintaining the good single-mode properties in the nanosecond regime. The key technological points are to arrange many nonlinear optical crystals in a layout to compensate for the walk-off arising from the angular phase matching and to precisely adjust the crystal spacing (typically with an accuracy of 0.1 mm), where the crystals are divided into pieces with the longest length to preserve the good single-mode properties (the walk-off should be kept within 1/10 for the fundamental beam diameter at $1/e^2$). As a typical example, we demonstrated a near-full energy conversion (efficiency 82%, output 8.7 mJ) while maintaining good single mode properties in the second harmonic (375 nm) generation process, in which we employed a single-frequency tunable nanosecond pulsed laser at a wavelength of 750 nm (output 10.6 mJ) as the fundamental.

The design principle presented here can be applied to other nonlinear wavelength conversion processes in the nanosecond regime. Furthermore, by extending this principle to incorporate a mechanism that dynamically manipulates the phase relationships between the crystals, it may be possible to create a device that can rapidly vary the wavelength-converted energy over the full dynamic range while keeping the total energy constant and the single-mode properties of the fundamental laser.

**Acknowledgements**
This work was supported by: Grant-in-Aid for Scientific Research (S) No. 20H05642, (A) No. 24244065, (B) No. 20H01837, JST PRESTO JPMJPR2105.